# Title: Thermal Stability Enhancement in Epitaxial α-Sn Films by Strain Engineering

Authors: *Huanhuan Song, Jinshan Yao, Yuanfeng Ding, Yu Gu, Yu Deng, Ming-Hui Lu, Hong Lu\*, and Yan-Feng Chen*

H. Song, J. Yao, Y. Ding, Y. Gu, Prof. M. Lu, Prof. H. Lu, Prof. Y. Chen
National Laboratory of Solid State Microstructures
College of Engineering and Applied Sciences
Nanjing University
Najing 210093, P. R. China
E-mail: hlu@nju.edu.cn
Prof. Y. Deng
National Laboratory of Solid State MicrostructuresCenter of Modern Analysis
Nanjing University
Nanjing 210093, P. R. China

## Abstract

Exploring new topological materials with large topological nontrivial bandgaps and simple composition is attractive for both theoretical investigation and experimental realization. Recently alpha tin (α-Sn) has been predicted to be such a candidate and it can be tuned to be either a topological insulator or a Dirac semimetal by applying appropriate strain. However, free-standing α-Sn is only stable below 13.2 °C. In this study, a series of high-quality α-Sn films with different thicknesses have been successfully grown on InSb substrates by molecular beam epitaxy (MBE). Confirmed by both X-ray diffraction (XRD) and reciprocal space mapping (RSM), all the films remained fully strained up to 4000 Å, proving the strain effect from the substrate. Remarkably, the single-crystalline α phase can persist up to 170 °C for the

200 Å thick sample. The critical temperature where the α phase disappears decreases as the film thickness increases, showing the thermal stabilization can be engineered by varying the α-Sn thickness. A plastic flow model taking the work hardening into account is introduced to explain this dependence, assuming the strain relaxation and the phase transition occur successively. This enhanced thermal stability is prerequisite for above room-temperature characterization and application of this material system.

The diamond-structured allotrope of tin, α-Sn, has attracted increasing research interest recently for its topological characters when the cubic symmetry is broken.[1-6] The advantages of this material are attributed to its simple structure and large tunable nontrivial bandgap.[2-12] However, α-Sn is a metastable phase that can transform to β phase easily above the phase transition temperature of 13.2 °C. Therefore most of the studies on α-Sn were conducted at low temperatures.[13-20] Farrow et al. demonstrated the substrate-stabilized epitaxial α-Sn films with elevated phase transition temperatures on InSb and CdTe substrates for the first time.[21] and similar results have been reported using different methods.[22-24] However, higher operating temperatures are still desired for practical application of α-Sn, e.g. spin-current convertor.[25] The crystal structure and related properties of α-Sn need further investigation, especially at an elevated temperature. This usually imposes a limitation on the film thickness of α-Sn from several to tens of monolayers (MLs).[2,5,8,12,25-27] Stanene, namely a monolayer of α-Sn, is one of the most popular 2D materials these years as it is predicted to be a large gap topological insulator [7,9-11,28,29] and related experimental results have been reported [30-36]. However, both the growth of stanene and the following device fabrication require extra care despite of its slightly higher thermal stability. We are interested in bulk α-Sn films for the reasons that thicker α-Sn films support more 3D topological phases[4-6], and benefit device fabrication. In this study, we focus on utilizing the substrate stabilization effect to systematically engineer the thermal stability of α-Sn films with thicknesses up to thousands of MLs.

In this work, we have grown a series of α-Sn films from 100 Å to 4000 Å on InSb substrate by MBE. The single crystalline α phase and the high interface quality are confirmed by transmission electron microscope (TEM) and XRD. RSM is used to examine the strain state of the films. At least up to the thickest film (4000 Å), there is no sign of relaxation at room temperature. The phase transition is investigated by both temperature dependent XRD and Raman spectra. We have observed a sudden disappearance of the α-Sn signals and report the transition temperature of 120 °C measured with XRD and 170 °C measured with Raman for 200 Å α-Sn sample. To explain the mechanism of the substrate-stabilization effect, we relate the phase transition to the strain relaxation of the films. The calculated self-stress of the dislocation at the interface is well consistent with the thickness dependent experimental results.

InSb(001) is chosen as the substrate in this study for its similar crystal structure to α-Sn. The α-Sn sample structure growth took place in two MBE systems, one III-V system (Veeco GENxplor) and one IV system (Dr. Eberl MBE-Komponenten). Firstly, the InSb substrate was introduced into the III-V system. After the oxide desorption of the substrate, an InSb buffer layer was grown to improve the surface smoothness. A typical (2×4) surface reconstruction can be observed by reflection high energy electron diffraction (RHEED). Then the wafer was took out of the vacuum and transferred to the IV system for Sn growth. An amorphous Sb capping layer was used to prevent surface oxidation and contamination, and later was removed in the IV system prior to the Sn growth. The sample structures are shown in **Figure 1**a. The Sn layer was grown at room temperature or below with a (2×2) surface reconstruction during the growth. More details of the sample growth can be found in the Experimental Section and **Table 1**.

Figure 1b shows the high-resolution XRD (Bruker D8) of the α-Sn samples with different thicknesses. The (004) diffraction peak of α-Sn can be clearly seen in the spectra especially when the sample is thicker than 50 nm. There are clear fringes that have never been reported, indicating smooth surfaces and interfaces. These fringes can be used to calculate the thicknesses of the films and the results agree well with the nominal values, also listed in Table 1. If we fix the lattice constant of InSb to be 6.480 Å, the out-of-plane lattice constant of the α-Sn film is calculated to be 6.499 Å, which is close to the value of unstrained α-Sn [6.489(2) Å].[37] The slightly higher value is expected due to the in-plane compressive strain (~0.14%). According to Hornstra and Bartels' approach,[38] we calculated the Poisson's ratio of α-Sn to be 0.35, consistent with that calculated from the elastic constants.[39] **Figure 2** shows the cross-sectional TEM images of a 200 nm α-Sn sample. The Sn/InSb interface in Figure 2a is hardly identified attributed to the similar cubic structures and atomic weights between α-Sn and InSb. The inset is the selected area electron diffraction (SAED) of α-Sn that confirms the single crystalline α phase and the coherent growth along [001] direction. The atomic resolution scanning TEM (STEM) images of α-Sn and InSb are shown in Figure 2b and 2c, respectively, confirming the desired crystal structures.

To investigate the strain effect and thermal stability of the α-Sn films, all the films were measured at room temperature on an asymmetric plane first. The RSM on (115) plane of the 400 nm thick sample in **Figure 3**a gives a typical example of a fully-strained epitaxial layer on the substrate. All the (115) peak positions of the samples in this study are summarized and shown in Figure 3b, compared to the theoretical line calculated using a Poisson's ratio of 0.298. The small deviation may come from the measurement resolution and the elastic deformation of the film in reality away from the interface. The calculated Possion's ratios using these peak positions are also listed in Table 1, agreeing well with the theoretical value.

All the aforementioned measurements were conducted at room temperature. It concludes that the α-Sn films are stable in ambient environment. Further investigations on the thermal stability were performed by temperature dependent XRD and Raman measurements and the results are shown in **Figure 4**. The sample was heated by 5 °C/min from room temperature to an elevated temperature. From the temperature dependent XRD (T-XRD) spectra (Figure 4a) of the 200 nm thick sample, as an example, a sudden reduction of the α-Sn diffraction peak intensity is observed at 70 °C. Simutaneouly, the (101) signals of β-Sn appear (see the Supporting Information). Figure 4b shows the temperature dependent Raman (T-Raman) spectra measured on the 400 nm α-Sn sample. The Raman shift peak of α-Sn (~197 cm$^{-1}$) can be clearly seen below 70 °C. Similarly, the disappearance of the α-Sn signal around 75 °C indicates the phase transition. We define the critical temperature as when the α-Sn signal disappeared.–Notably, the α-Sn signal never appeared again during the cooling process in both measurements. This result is different from those of Farrow and Menéndez in which the phase transitions were partially reversible.[21,23] Qualitatively, the two methods used to define the critical temperature reflect complimentary aspects of the sample thermal stability. XRD reveals the single crystalline quality, especially the interface quality of the epitaxial layers, through the interference fringes, while the Raman signals are believed mainly from the topmost layers. In this study, due to the narrow bandgap and strong absorption of α-Sn, the signal of the InSb substrate is too weak to detect. All the critical temperatures are plotted in Figure 4c as a function of the film thickness. The critical temperature decreases as the thickness increases, and saturates to about 70 °C when the thickness is above 200 nm. Another characteristic feature in Figure 4a may provide an insight into the phase transition of α-Sn films. The fringes started to disappear as the temperature approached the critical temperature, indicating the degradation of the interface, i.e. formation of considerable dislocations at the interface during the film relaxation. This may occur right before, or at the same time with, the phase transition of α-Sn. In order to explain the increased critical temperature and its thickness dependence, we introduce a model to describe the relaxation process, where both thermal dynamic and kinetic conditions are considered.

The equilibrium between two phases is due to the competition between two forces, the driving force $\Delta G_{dr}$ and the resistance $\Delta G_{re}$ (or barrier). So the free energy change is what matters. A typical transformation of bulk α-Sn to β-Sn is a vibrational entropy-driven process.[40,41] However, the epitaxial α-Sn film undergoes a more complicated process because of its coherent relationship with the substrate. In this case, the elastic deformation due to the lattice mismatch and the direct bonding with the substrate are two most important factors. The strain relaxation will change the energy levels of these two factors dramatically. Here, we use the plastic flow model proposed for the strained heterostructures relaxation to qualitatively analyze the strain relief process of the α-Sn films,[42,43] during which the driving force of the phase transition should be given a correction. The process started with a strained α-Sn epilayer, ended with a relaxed layer in another phase, and underwent the breaking of the interfacial bonds, relief of the mismatch strain and phase transition in between. So the total free energy change is expressed as follows

$$\Delta G_{free} = \Delta G_{pt} + \Delta E_{elastic} + \Delta E_{df} + \Delta E_{dm} \qquad (1)$$

where $\Delta G_{pt}$ is the free energy change of bulk α-Sn phase transition; $\Delta E_{elastic}$ is the released elastic energy; $\Delta E_{df}$ and $\Delta E_{dm}$ correspond to the energy of the dislocation formation and dislocation motion, respectively, more crucial in the plastic flow model. Here we neglect the entropy change in the elastic deformation and dislocation formation and movement. Specifically, at room temperature $\Delta G_{pt}$ and $\Delta E_{elastic}$ should be negative, so they are classified as $\Delta G_{dr}$, while $\Delta E_{df}$ and $\Delta E_{dm}$ provide the barrier $\Delta G_{re}$ for strain relaxation because it involves breaking the interfacial bonding and induces local deformation.

Obviously, if $\Delta E_{elastic} + \Delta E_{df} + \Delta E_{dm} > 0$, even at the normal transition temperature, there will be a net barrier for phase transition, so the phase is still stable. This is true when the film thickness is below the critical thickness. As temperature increases, $\Delta G_{pt}$, $\Delta E_{df}$ and $\Delta E_{dm}$ all decrease and eventually the phase transition condition $\Delta G_{free} < 0$ is reached.

According to the theory of Fischer et al.,[44,45] the effective shear stress that acts on a misfit dislocation in a slip plane can be expressed as

$$\tau_{eff} = \tau - \tau_{pf} - \tau_s \qquad (2)$$

with the resolved shear stress $\tau$ of in-plane stress due to internal strain in a pseudomorphic epilayer

$$\tau \propto \frac{2G(1+\nu)}{1-\nu} \qquad (3)$$

the strain relief via plastic flow being expressed as

$$\tau_{pf} \propto \frac{2G(1+\nu)}{1-\nu} \cdot \frac{b\cos\lambda}{p} \qquad (4)$$

and the shear component of the self-stress due to the elastic dislocation interactions

$$\tau_s \propto \frac{Gb(1-\nu\cos^2\theta)}{4\pi R_{h,p}(1-\nu)\cos\lambda} \ln\left(\frac{\alpha R_{h,p}}{b}\right) \qquad (5)$$

where the in-plane strain $\epsilon = (a_l - a_s)/a_s$, $a$ denotes the lattice constant and the subscript $l$ and $s$ refer to the epilayer and the substrate, respectively; $G$ and $\nu$ are the shear modulus and Poission's ratio of the epilayer, respectively; $b$ is the magnitude of the Burger's vector; $p$ is the average distance between the dislocations. $\alpha$ is a factor accounting for the energy in the dislocation core, typically 1 to 4 for covalently bonded semiconductors; $\theta$ is the angle between the Burger's vector and the dislocation line and $\lambda$ is the angle between the Burger's vector and the normal of the dislocation line in the interface. The complex radius $R_{h,p}$ involves both the film thickness $h$ and the average distance between the dislaocations at the interface when the boundary condition of free surface is considered. During the initial stage of strain relaxation with low dislocation density, $h \ll p$, so $R_{h,p} \approx h$; otherwise $R_{h,p}$ approaches $p/2$ when $h \gg p$ as a result of either dislocation multiplication or thickness increment.

Applying the parameters of α-Sn into Equation (5), $G$ =25 GPa, $b$ =4.59Å, $v$ =0.298, $cos\lambda = cos\theta$ =0.5, we calculated $\tau_s$ in the initial stage as a function of thickness as shown in Figure 4c. It shows a similar thickness dependence as the critical temperature. Usually $\Delta E_{elastic}$ increases linearly with the layer thickness and serves to lower the transition temperature, however, it cannot explain the saturation with lager thickness. So the strain relaxation process in α-Sn films is probably dominated by $\Delta E_{dm}$, which is closely related to the self-stress. It is the effective shear stress in the slip plane $\tau_{eff}$ that provides the gliding force. So the interactions between dislocations should be the main resistance which is actually determined by $\tau_s$. This is the analogy to the work hardening effect in materials. For the film with larger thickness, though $\tau_s$ in the initial stage gives a similar saturation behavior, however, the $h \ll p$ condition may fails during the strain-relaxation process, and finally $R_{h,p} \approx p/2$. Anyway, the phase transition temperature will become thickness independent with increasing thickness.

The validness of the above discussion can be inferred from a kinetic description

as well. Brian et al. proposed that time-dependent relaxation via plastic flow is[43]

$$\frac{d\epsilon_{pf}(t)}{dt} = CG^2\bigl[\epsilon - \epsilon_{pf}(t) - \epsilon_s(h)\bigr]^2 \epsilon_{pf}(t) \tag{6}$$

where the strain terms $\epsilon_{pf}(t)$ and $\epsilon_s(h)$ correspond to the stress terms $\tau_{pf}$ and $\tau_s$; $C$ is a temperature dependent factor and increases exponentially with temperature. This relation points out that at initial stage the strain-relaxation rate is very small due to the small $\epsilon_{pf}(t)$. The increasement of $\epsilon_{pf}(t)$ will accelerate the relaxation and finally settle due to the vanishing effective shear stress. The residual strain in the film depends on $\epsilon_s(h)$. Brian et al. and Fischer et al. used their models to explain the much larger critical thicknesses of SiGe alloys observed in experiments.[43-45] In this study the much higher critical temperature than usual phase transition temperature can be interpreted in the same way. On the other hand, the hysteresis character again emphasize that it is the initial stage limiting the strain-relaxation process. So the assumption $R_{h,p} \approx h$ for comparison with the thickness dependent behavior is reasonable.

The exact timescale depends on the degree of strain relief. In this study, we may assume a critical strain relief $\epsilon_{pf}^c$ which is responsible for the disappearance of α-Sn. Now the dynamic condition to observe the disappearance of α-Sn is the time needed to reach this critical value $t_c$ is within the measurement timescale $\Delta t_m$. According to Equation (7), the rate of strain relief increases with $\tau_{eff}$ and temperature. Therefore the larger the thickness is, the shorter the $t_c$ is, and the lower the critical temperature is. The schematic dependence of $t_c$ on thickness and temperature is shown in the inset in Figure 4c. The exact value of $\epsilon_{pf}^c$ is not critical to affect this dependence.

Therefore, the critical temperatures measured by T-Raman are generally higher than those measured by T-XRD because the latter takes much longer time for temperature stabilization as well as each measurement.

In conclusion, we have successfully grown a series of single crystalline α-Sn films on InSb substrates and systematically investigated their thickness dependent thermal stability. The highest critical temperature is over 120 °C for the 20 nm sample. The high-quality interfaces for all the samples have been confirmed which have been used as the critical factor in the plastic flow model. Any initial defects at the interface may facilitate the phase transition through dislocation multiplication. The good agreement between the phase transition temperature and $\tau_s$ indicates the elastic interaction between the dislocations is the dominant factor for thickness dependent behavior. Therefore we can construct a phase diagram predicting the thermal stability of epitaxial α-Sn films on InSb substrates. There is a divergence for very thin α-Sn films. The extremely high thermal stability of very thin films is favorable for higher growth temperatures and various experimental treatments, post-growth annealing for example.[30,34-36] Our results provide an important evidence for reliable measurements on α-Sn and potential applications above room temperature.[25,27]

**Acknowledgements**
H. S., J. Y., and Y. D. contributed equally to this work. The authors acknowledge the support from the National Key R&D Program of China (2018YFA0306200, 2017YFA0303702), the National Nature Science Foundation of China (Grant No.51732006, No.11890702, No.51721001), the Thousand Talents Program and Jiangsu Entrepreneurship and Innovation Program.

**Supporting Information**

Supporting Information is available from the Wiley Online Library or from the author.

**Experimental Section**

*Sample Growth*: the growth of the α-Sn samples went through a two-step process. Firstly, the oxide desorption of the InSb substrate was performed thermally up to 500 °C with an $Sb_2$ overpressure of $5\times10^{-7}$ torr in the III-V MBE system (Veeco GENxplor$^{TM}$). After the (2×4) surface reconstruction was observed by RHEED, the substrate temperature was decreased to 450 °C, at which an InSb buffer layer was grown on the InSb substrate using an indium flux of $5.4\times10^{-8}$ Torr for about 30 mins. Then the substrate temperature was lowered to 100 °C. An amorphous Sb capping layer was deposited on the InSb surface for at least 30 mins to prevent the oxidation during the template transfer. The template was transferred to the IV MBE system (Dr. Eberl MBE-Komponenten Octoplus 300) for the second step growth. After being transferred into the IV MBE system, the substrate temperature was raised to 400 °C, at which the Sb capping layer was removed by evaporation until the typical (2×4) surface reconstruction of InSb appeared again. Then the substrate temperature was lowered to below room temperature for Sn growth. The Sn film was grown on the InSb surface by heating Sn (99.9999 %) effusion cell to 1200~1300 °C for an appropriate growth rate. The background vacuum level was about $1\times10^{-9}$ torr.

*Temperature Dependent Measurement*: To investigate the strain status and the thermal stability of the α-Sn films, the RSM and the temperature dependent XRD were measured on a Bruker D8 Discover high resolution XRD instrument. We scanned the full spectra at room temperature. The temperature dependent XRD was measured using a ramp rate of 5 °C·min$^{-1}$ with an interval of 5 or 10 °C. Before each measurement there was a 5~10 min settling time to stabilize the temperature, and each scan took about 15 mins. The temperature dependent Raman was measured using a 488 nm Ar$^+$ laser on a Horiba LabRAM HR Evolution Raman spectrometer. The temperature ramp rate is 5 °C·min$^{-1}$ and the interval is 5 or 10 °C. Before each measurement there was a 2 min settling time for temperature stabilization, and each scan took about 2 mins.

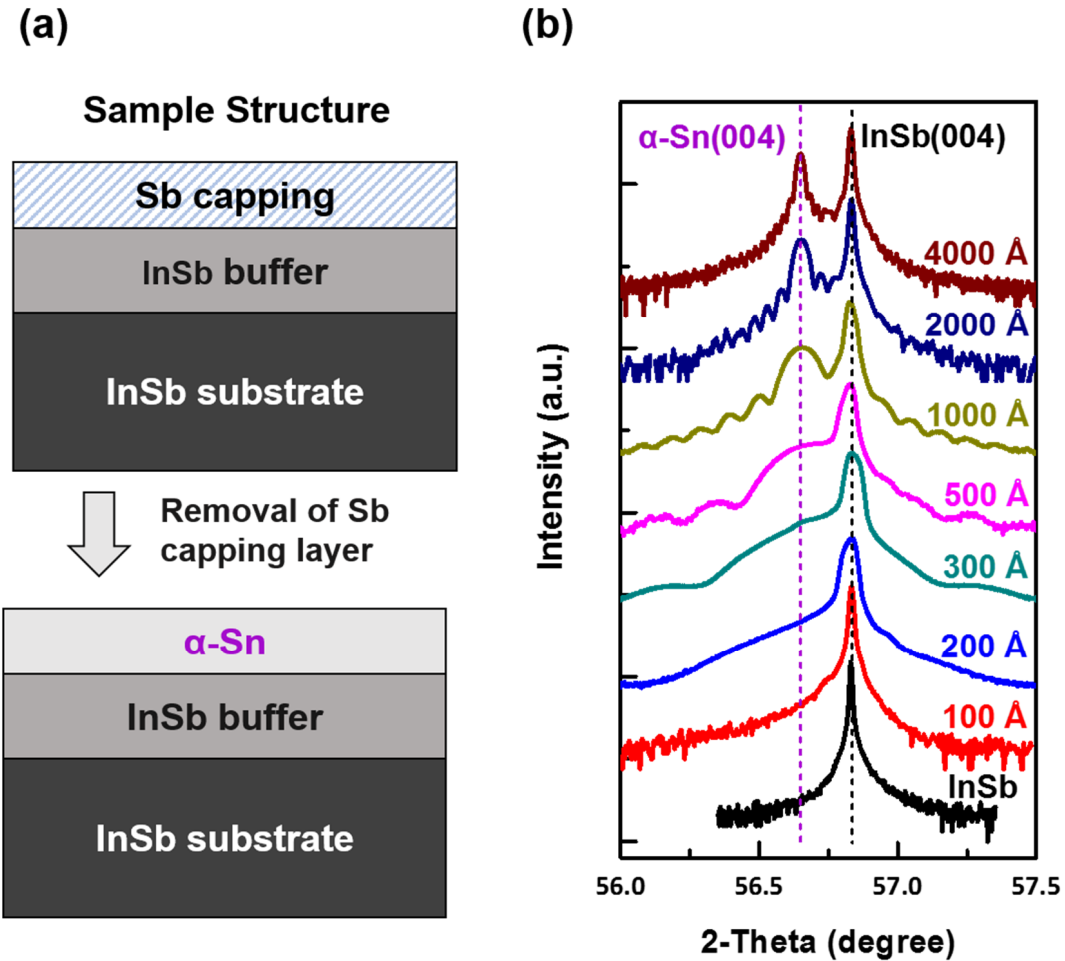

**Figure 1.** Sample structures of α-Sn films grown on InSb substrates. a) The schematic layer structures of the α-Sn/InSb heterostructure showing the two-step growth procedure. The upper structure shows the InSb buffer and Sb capping layer grown in the III-V MBE system. The lower structure shows the α-Sn layer growth after removal of the Sb capping layer in the IV system. b) High-resolution X-ray diffraction spectra of α-Sn/InSb samples with different thicknesses. The (004) diffraction peaks of α-Sn and InSb are marked with purple and black dashed lines, respectively. An InSb substrate spectrum is also shown as a reference.

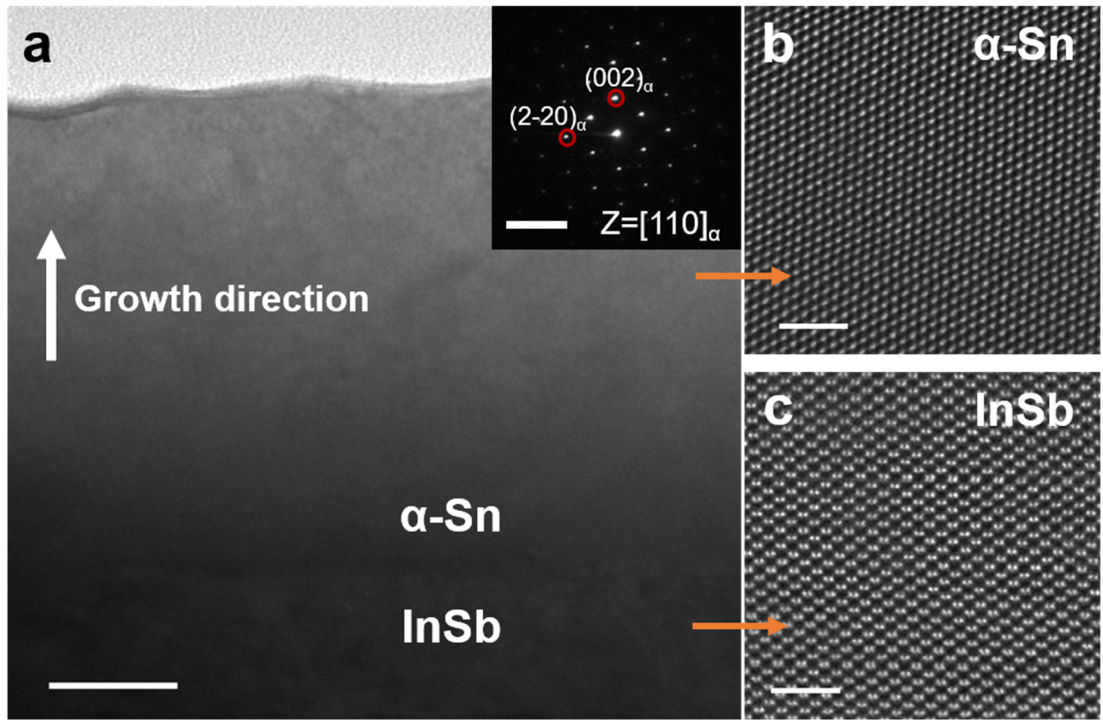

**Figure 2.** TEM characterization of the α-Sn films. a) The cross-sectional TEM image of an α-Sn/InSb heterostructure. The scale bar is 50 nm. The inset is the SAED pattern of the film region. The scale bar is 0.2 nm$^{-1}$. b) and c) are the STEM images of α-Sn and InSb, respectively, with atomic resolution. Both scale bars are 2 nm.

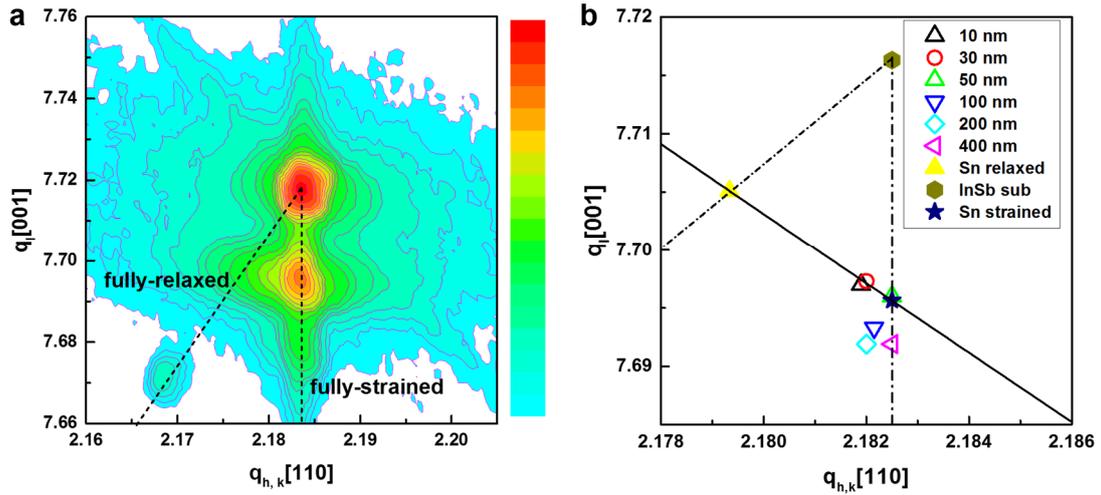

**Figure 3.** The asymmetric (115) RSM on the α-Sn films. a) The RSM of a 200 nm α-Sn/InSb sample. (b) The α-Sn peak positions obtained from the RSM of all the α-Sn samples with different thicknesses. The dashed lines correspond to the fully-strained and fully-relaxed conditions. The black solid line is the theoretical prediction calculated from $a_L = (1-v)a_\perp/(1+v) + 2va_\parallel/(1+v)$ with Poisson's ratio $v$ of 0.298.

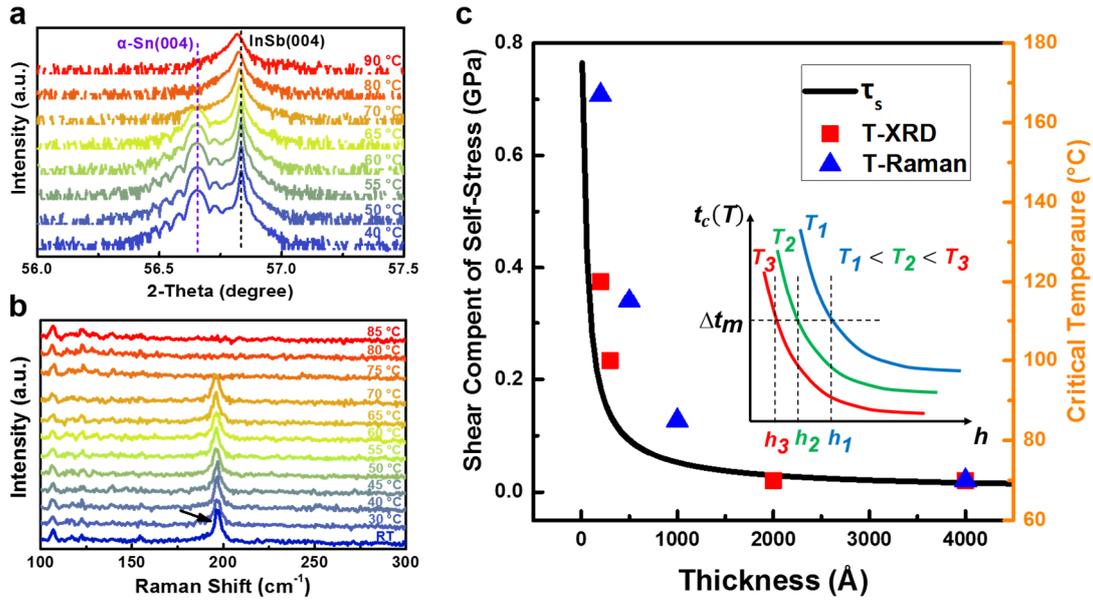

**Figure 4.** Thermal stability investigations on the α-Sn films. a) The temperature dependent XRD of a 200 nm α-Sn film. The (004) diffraction peaks of α-Sn and InSb are marked with purple and black dashed lines, respectively. The α-Sn diffraction peak disappears above 70 °C. b) Temperature dependent Raman spectra of a 400 nm α-Sn film. The Raman shift peak of α-Sn is pointed by the black arrow and disappears above 70 °C. c) The critical transition temperatures of the α-Sn films obtained from temperature dependent XRD and Raman spectra as a function of thickness. The red square and the blue triangle represent the data from XRD and Raman, respectively. The black solid line is the calculated self-stress of dislocations. The inset is the schematic thickness dependence of $t_c$ at three different temperatures, $T_1 < T_2 < T_3$. $\Delta t_m$ is the measurement timescale.

**Table 1.** The summary of relative parameters including nominal thickness, actual thickness, growth rate, growth temperature and Poission's ratio of the α-Sn films in this study.

| Sample | Nominal Thickness[a] [nm] | Actual Thickness[b] [nm] | Growth Rate [Å/s] | Growth Temperature [°C] | Poission's Ratio[c] |
|---|---|---|---|---|---|
| A | 10 | 9.3 | 0.025 | 10 | 0.309 |
| B | 20 | 18.6 | 0.025 | 10 | 0.136 |
| C | 30 | 27.9 | 0.025 | 14 | 0.293 |
| D | 50 | 46.5 | 0.025 | 16 | 0.290 |
| E | 100 | 96.5 | 0.085 | 17 | 0.373 |
| F | 200 | 206.1 | 0.08 | 14 | 0.413 |
| G | 400 | 407.1 | 0.28 | 15 | 0.372 |

[a] Obtained from QCM measurements; [b] Simulated from XRD data; [c] Calculated using equation $a_L = (1-v)a_\perp/(1+v) + 2va_\parallel/(1+v)$ with the lattice constant of the relaxed layer $a_L$, the out-of-plane lattice constant $a_\perp$ and the in-plane lattice constant $a_\parallel$. The in-plane and out-of-plane lattice constants are derived from the asymmetric (115) RSM.